# A Review on the effectiveness of Dimensional Reduction with Computational Forensics: An Application on Malware Analysis

**Aye Thaw Da Naing, Justin Soh Beng Guan, Yarzar Shwe Win, Jonathan Pan**

Nanyang Technological University, Singapore[1]

Home Team Science and Technology Agency, Singapore[2]

ayethawd001@e.ntu.edu.sg[1], S210054@e.ntu.edu.sg[1], YARZAR002@e.ntu.edu.sg[1],

Jonathan_Pan@htx.gov.sg[1,2]

## 1. ABSTRACT

The Android operating system is pervasively adopted as the operating system platform of choice for smart devices like smartphones, tablets, home appliances and Internet of Things (IoTs). However, the strong adoption has also resulted in exponential growth in the number of Android based malicious software or malware. Such malwares typically embed themselves in their victims' devices and attack not only their victims but induce other targeted or collateral damages. To deal with such cyber threats as part of cyber investigation and digital forensics, computational techniques in the form of machine learning algorithms are applied for such malware identification, detection and forensics analysis. However, such Computational Forensics modelling techniques are constrained the volume, velocity, variety and veracity of the malware landscape. This in turn would affect its identification and detection effectiveness. Such consequence would inherently induce the question of sustainability with such solution approach. One approach to optimise effectiveness is to apply dimensional reduction techniques like Principal Component Analysis with the intent to enhance algorithmic performance. In this paper, we evaluate the effectiveness of the application of Principle Component Analysis on Computational Forensics task of detecting Android based malware. We applied our research hypothesis to three different datasets with different machine learning algorithms. Our research result showed that the dimensionally reduced dataset would result in a measure of degradation in accuracy performance.

## 2. AUTHOR KEYWORDS

Principal Component Analysis (PCA), Computational Forensics, Android Malware.

## 3. INTRODUCTION

The Android operating system (OS) continues to dominate the market share of mobile devices around the world. Android OS has widely been used in automotive, IoT (Internet of Things) devices, home appliances and smart watch. With Android powered mobile devices, they have enabled users to access to internet-based communications, emails and social media without the need for computers (Kalkbrenner, J., et al 2011). The integration of mobile payment capabilities into smartphones provide users with digital mobile wallets and contactless payment (Bezovski, Zlatko et al., 2016, Slade, Emmaet al., 2013). However, the threat of malware to Android OS has been growing over the past 10 years (Feizollah, Ali et al., 2017). To deal with epidemiological spread of Android based malware, Google developed Google Play Protect to secure and scan all mobile app submissions for embedded malware (Sawers, P., 2020). However, despite the preemptive step to contain the malware spread, Android platforms are still exposed to malware infiltrations and infections. Hence to contain this cyber epidemiological disaster, it is crucial for cyber investigators and digital forensics analysts using machine learning based classifiers have an effective means to deal with voluminous, veracity and variety of malware.

Most existing android malware detection system and frameworks can be categorized into three groups, namely, Static analysis, Dynamic analysis, follow by a hybrid of the two methods. Static analysis detects Malware through source code permission and intent which allows fast detection. Prior to installation, the APK application is dissected, with its content such as Android Manifest.xml and DEX (Decentralized Exchanges) files being analyzed to determine if it's malicious. However, modern Android malware employs code obfuscation techniques to evade static analysis (Abdullah Talha Kabakus et al., 2018). Dynamic analysis investigates the actual behavior and processes of suspicious application in a real time environment to detect the presence of malware or malicious code. Dynamic analysis requires execution of APK on emulator or physical devices, refer to as sandbox, requiring sizeable amount of processing power and time (Elsersy, Wael et al., 2022). The use of hybrid analysis is becoming common in recent years. Many frameworks which combine static and dynamic analysis to characterize the behavior of malware analysis. (Abdullah Talha Kabakus et al, 2018, L. Taheri et al., 2019). Some researchers make use of Machine Learning algorithms to identified Android malware from benign software based on features from static, dynamic or hybrid analysis (Meghna Dhalaria et al, 2020). New methods of analysis which converts dissected APK files into datasets to perform classification have been deployed to improve Malware classification (Y. Fang et al.,.2020).

Moreover, more and more studies have been done to find a way to improve Machine Learning models. One of the methods is utilizing dimensional reducing methods such as Principal Component Analysis (PCA), Linear discriminant analysis (LDA) or T-Distributed Stochastic Neighbour Embedding (T-SNE). In our study, we evaluate the application of dimensional reduction method namely PCA and evaluate the accuracy performance of machine learning classifier algorithms to detect Android malware. In the next section, we will cover the related literature to our research. This is followed by a description of the research experiment that we applied that included the datasets involved and experimentation steps taken. An analysis of our research results follows. This is then concluded with our conclusion.

## 4. LITERATURE REVIEW

There has been a growing number of android malwares. According to Statista Research, as of month of March 2020, a total of 482,579 android malware have been detected. (statista.com.,. 2022) There are various research done in the identification of android malware based on static and dynamic analysis.

Research on the effectiveness of dynamic analysis of Android intent features and Android Permission features in Android malware detection had a detection rate of 91% and 83% respectively, with a combination of both intent and permission features achieving a higher detection rate of 95.5% (Feizollah, Ali et al., 2017). Earlier research proposed on the use of static analysis based solely on permissions and creating probabilistic generative model for risk scoring (Peng, et al.,.2012). Stowaway, a tool developed to detect over privilege of Application Programming Interface (API) calls and mapping these set of API calls to permissions (Felt, Adrienne et al., 2011). Information on the permission required of an android application can be found in Androidmanifest.xml file in the apk which can be extracted using AXMLPrinter2 tool (P. P. K. Chan et al., 2014). Unlike static analysis, which is vulnerable to code obfuscation, dynamic analysis monitors the artifacts generated by the executed apk in physical phone or virtual environment. (Feizollah, Ali et al., 2017). Research have been conducted using CICAndMal2017 dataset to generate network traffic on actual smartphones using a systematic approach rather than virtual emulators (Habibi Lashkarii et al., 2018).

Machine learning (ML) classifiers such as Decision Tree (DT), Random Forest(RF), K-Nearest Neighbors(KNN), Naives Bayes(NB) and Support Vector Classifier(SVM) are common supervised learning algorithms used by researchers to perform both binary and family classification of malware (Noorbehbahani et al., 2019, Dhalaria, M et al.,. 2020, Sangal, Aviral et al., 2020, Abdullah, Talal et al.,. 2020). Research on evaluating the performance of permission feature dataset compared to permission and API calls dataset used common ML classifiers that includes

Naïve Bayes, Support Vector Machine (SVM), decision tree and Random Forest (RF) (P. P. K. Chan et al., 2014, S. E. Mohamed et al., 2021). These traditional ML classifiers are often used by researchers as baseline to compare against the performance of deep learning models and framework (M. Masum et al., 2019, El Fiky, A. H., 2020). Common evaluation metrics used by researchers include Accuracy, F1, Precision, Recall, True Positive Rate (TPR), False Positive Rate (FPR) and Area under Curve(AUC). Researcher Samaneh Mahdavifar used semi-supervised deep neural network algorithm to perform category classification of malware. In his work, he created the CICMalDroid2020 dataset consisting of five categories of Android malware, namely, Adware, Banking, SMS (Short Message Service), Riskware and Benign. Common tools such as CuckooDroid and CopperDroid are commonly used to collect dynamic dataset of APK (Mahdavifar, Samaneh et al., 2020, Dhalaria, M et al., 2020). Aside from ML, Natural language processing algorithms were used to extract ASCSII strings from Android APK. These were further processed into individualized words. With these collection of words, they were then converted into lexical features. These features vectors are then used as inputs into ML classifiers such as random forest and convolution neural network. The combination of these techniques were assessed to be effective in the detection of Android malware (Mimura, M., 2022).

Feature selection techniques have typically used to reduce the features that are not useful in the dataset to improve accuracy of the ML models (Fiky A. H. E., et al, 2021). Such techniques also improve the processing time and prevent model over-fitting hence resulting with models that are more robust and generalized. Common feature selection technique includes Information Gain (IG) technique which ranks a feature by calculating the information gain. The need to process large volume of data have also result in dimension reduction techniques gaining much of the attention. Testing of ML classifiers with large datasets can be time consuming. Dimensionality reduction reduces the high dimensional vector-valued explanatory variables while able to preserves its relationship with a low dimensional space (Zhang, T et al., 2018). Research was done to explore other ways to improve the processing time. This includes training ML classifiers with reduced dataset of smaller sizes based on random sampling and stratified random sampling. There has been some research on the application of PCA on analysis of Android malware (D, Arivudainambi et al., 2019). Studies have also been conducted to compare the performance difference of dimension reduction techniques with PCA and LDA (Durmuş Özkan Şahin et al., 2021). Most of these studies do not make direct reference to experimental results from other literature or to use different malware datasets to understand the benefits of PCA. Combination of PCA and feature selection technique IG have been studied by researcher El Fiky (El Fiky, A. H., 2020) to create an optimized technique to reduce 89% of total features. They tested with three baseline classifiers and managed to achieve good F-measure results with Random Forest. However, the combination of Drebin and Malgenome dataset used in the research does not address to the issue of imbalance nature of both datasets. The entire Drebin dataset contains 4.3% of malware (Xu, Jiayun.,. et al, 2021). Both datasets have often been used by various researchers to develop frameworks to improve malware classification and detection (M. Masum et al., 2019, El Fiky, A. H., 2020). There is a need to investigate the impact of unbalance dataset on Android malware detection research. More studies are needed to investigate the effects of dimensionality reduction technique, feature selection techniques and balancing of datasets to improve on accuracy of malware detection hence the relevance of this research work.

## 5. PROPOSED METHODOLOGY

### 5.1 Datasets

The proposed methodology involves using three different datasets, namely, CICInvesAndMal2019, Drebin and Android Malware Detection and Classification for analysis and research.

Dataset-1 used in our research is Malgenome dataset. The dataset consists of 1260 Android malwares belonging to 49 malware families and 2539 benign APKs. The dataset has feature vectors of 215 attributes. The entire dataset took more than a year of reading through security blogs from existing anti-virus companies, lodging requests for samples and web crawling to obtain the malware samples (Y. Zhou et al., 2012). Due to limited resources, they have since stop updating the dataset from 2015 onwards. However, other researchers further processed the dataset collection to extract the features by decompiling Android manifest files using the tool AXMLprinter2. API calls were also extracted using Baksmali disassembler tool (S. Y. Yerima.,. 2019).

Dataset-2 used in our research is the Drebin dataset initially created by MobileSandbox project. (M. Spreitzenbarth., 2013) It includes 5,560 malwares from 179 different malware families. Drebin datasets include data on static analysis such as applications' manifest, dex code and permissions. (D. Arp., 2014) In our experiments, we utilized the dataset extract by S. Y. Yerima for their paper "DroidFusion: A Novel Multilevel Classifier Fusion Approach for Android Malware Detection" in 2019. This dataset includes 5,560 malwares and 9,476 benign apps. The dataset also uses 215 features and contains 2 classes to classified malware and benign.

Dataset-3 used in our research was called CIC-InvesAndMal2019 dataset (Sangal, Aviral., 2020). The dataset was retrieved from Canadian Institute for Cybersecurity. The dataset includes permissions and intents as static features and API calls. The dataset has 5,491 collected samples with 426 malware and 5,065 benign. There are four malware classification in the dataset. They are Adware, Ransomware, Scareware and SMS Malware. The following table summarises the datasets used with our research work.

| Datasets | Number of samples | Number of malwares | Number of benign | Number of features |
|---|---|---|---|---|
| Malgenome | 3799 | 1260 | 2539 | 215 |
| Drebin | 15036 | 5560 | 9476 | 215 |
| CIC-InvesAndMal2019 | 5,491 | 426 | 5,065 | 253 |

**Table 1: Datasets and their samples**

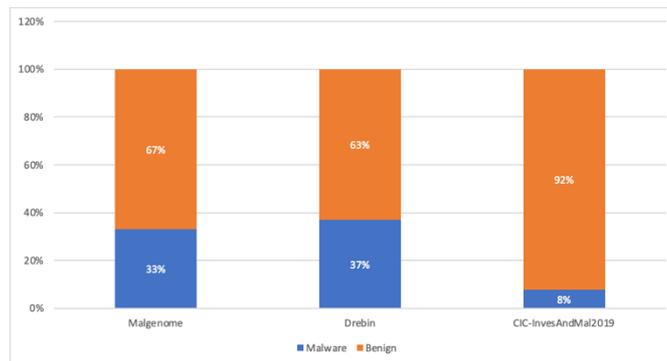

**Figure 1: Number of Malware and Benign Samples in Each Dataset**

## 5.2 Methodology

Our experiment involved that replication of the use of software tools or Python development packages along with machine learning parameters mentioned in cited literature. For those literature without information regarding detailed parameters, we selected and tested the best parameters based on nearest metrics.

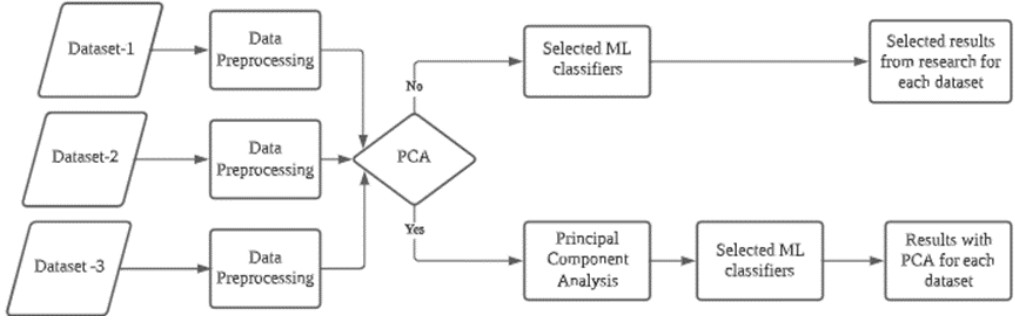

**Figure 2: Proposed Approach Illustration**

Based on the three different datasets, we compared the test results from cited literature with the results of dataset processed with PCA prior to splitting the dataset into train and test to be used on ML classifiers based on respective literature. Some of the research may contain specific steps in data processing which will be mentioned in experiment and results. We apply K-Folds cross validation to ensure less biasness when training and testing our model (Wazirali, Ret al., 2020). Most of the results are obtained with tenfold cross validation using WEKA and python. (S. Y. Yerima et al.,.2019, Akintola A.G. et al., 2022, El Fiky, A. H., 2020). Based on the evaluation metric, the performance of our testing of ML classifiers were compared with cited literature using similar dataset to validate if dimensionality reduction can be applied to Computation Forensic while achieving satisfactory results.

## 5.3 Performance Evaluation Metrics

Since various kinds of literature use different evaluation metrics, our study will be based on multiple different evaluation metrics in cited literature instead of using the common metrics.

i. Accuracy measures the overall rate at which the model correctly predicts the label:

$$Accuracy = \frac{TP + TN}{TP + FP + TN + FN}$$

ii. F-score is the harmonic mean of both the recall (R) and precision (P) metrics. It is commonly used to evaluate the performance of binary classification model. F-score can be computed as defined in Equation:

$$Fscore = \frac{2 \times TP}{2 \times TP + FP + FN}$$

F-score can be enhanced into F-beta score where beta is used to choose the weight between precision and recall.

$$F_\beta = (1 + \beta) \frac{Precision \times Recall}{(\beta^2 \times Precision) + Recall}$$

iii. Precision is used to measure the True Positive Rate of the dataset and can be calculated by using below Equation:

$$Precision = \frac{TP}{TP + FP}$$

iv. Recall is commonly used to measure how much of the dataset is accurately identified:

$$Recall = \frac{TP}{TP + FN}$$

# 6. RESULTS AND DISCUSSION

In this section, we identified 2 literatures for each dataset. To get comparable results for ML classifiers used in past literature, we attempt to reproduce the experiment scenario. The dataset is then processed with PCA and passed into ML classifiers. This is to ensure the reliability of our test results.

## 6.1 Results From Malgenome Dataset

We selected two relevant literatures that used on Malgenome dataset. The first literature is Empirical Analysis of Forest Penalizing Attribute and Its Enhanced Variations for Android Malware by Akintola A.G. which is a journal from MDPI (Akintola A.G. et al., 2022). The second literature we selected is Empirical Study on Intelligent Android Malware Detection based on Supervised Machine Learning by Abdullah T.A, published in IJACSA in United Kingdom (Abdullah Talha Kabakus et al., 2018). Malgenome dataset is commonly used in research on ML classifier performance, study on effectiveness of intents, permission and API calls in malware classification and application of dimensionality reduction techniques to achieve high performance in ML algorithms with less computational resources (S. E. Mohamed et al.,. 2021). (Refer to Appendix Section 1 for features breakdown).

In the first literature, Akintola A.G. conducted an empirical study to validate using Forest Penalizing Attribute (FPA) classifier, followed by enhanced FPA to detect android malware. In our study, we will not investigate into the enhanced FPA variants. Figure 3 shows the baseline classifiers used to compare with FPA (Akintola A.G. et al., 2022).

| Baseline Classifiers | |
|---|---|
| Naïve Baye | NB |
| Baye Net | BN |
| Conjutive Rule | CR |
| Decision table | DETAB |
| Alternating Decision Tree | ADT |
| Decision Stump | DS |

Figure 3: Baseline Classifiers used in Akintola research

Synthetic minority oversampling technique, known as SMOTE, was used by Akintola A.G. to solve class imbalance issues found in the Malgenome dataset. Our research plan was to apply PCA on the Malgenome dataset and evaluate the resultant dataset using the performance evaluation metrics. To reproduce her results, we used WEKA and performed a K-fold cross validation with k-fold set to 10 folds. Apart from +2% improvement in DETAB for accuracy over researcher Akintola A.G. results, most results are consistent. NB kernel estimator parameter is set to True. (Refer to appendix Section 1 for Table 1 Original dataset Results)

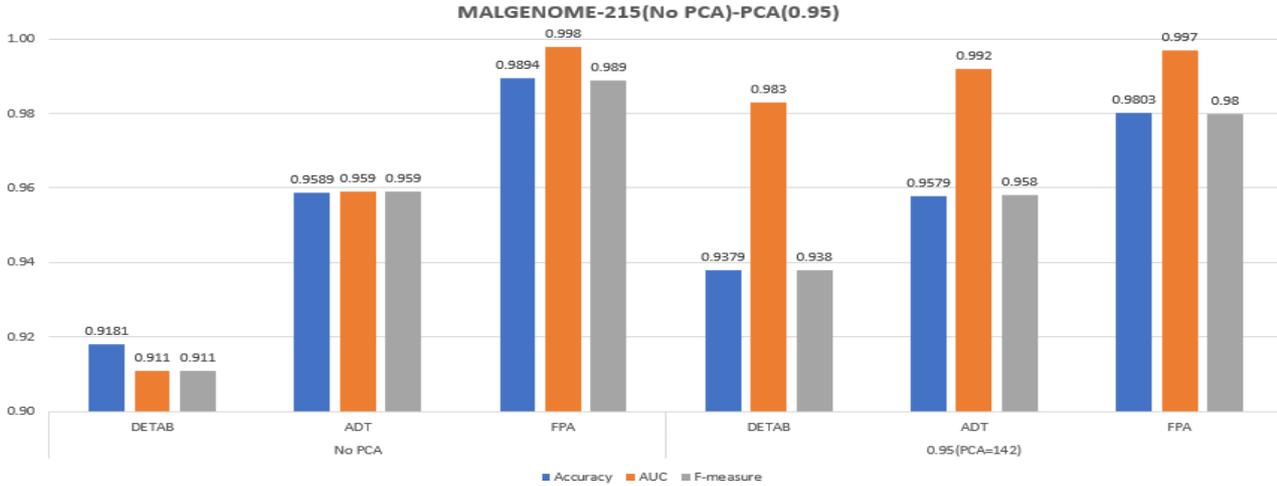

**Figure 4: Top three algorithms before and after PCA**

Figure 4 shows the top three algorithms before and after PCA. FPA performs the best among all the classifiers with highest accuracy at 0.9894. (Refer to appendix section 1 Table 2 for PCA results)

The accuracy increased by 0.002 for DETAB, a drop of 0.001 for ADT (Alternating Decision Tree) and 0.009 drop for FPA. We can see that PCA successfully reduced the number of features from 215 to 142 while continuing to achieve good results for these algorithms.

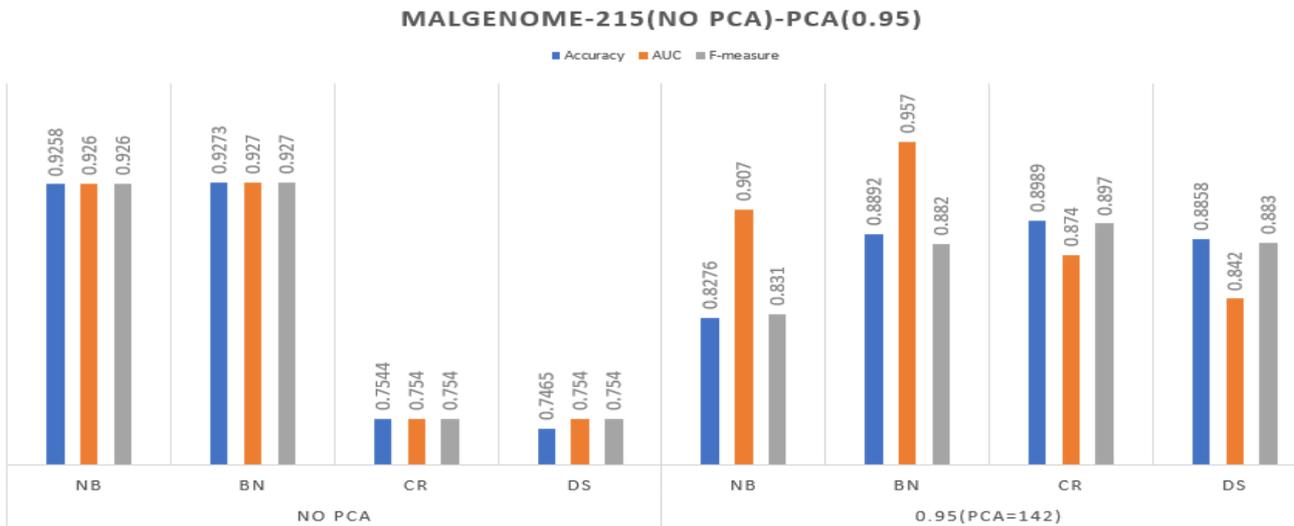

**Figure 5: Remaining algorithms before and after PCA**

In figure 5, we examine the algorithms with lower performance. While most classifiers have accuracy of above 0.9, both CR (Conjunctive Rule) and DS had accuracy measurements below 0.8 before PCA. When PCA with variance of 0.95 was applied, both NB and BN (Baye Net) suffered a drop of 0.098 and 0.038. However, DS and CR improved 0.140 and 0.145 respectively. Weak models such as DS and CR were unable to handle a large number of features in the dataset. As DS is a one level decision tree, the first few features of PCA managed to capture majority of the variance in the dataset (Wayne Iba et al., 1992, . J. Chandrasekaran.,. 2020). This gives the first feature of PCA more prediction power than the first feature in the original dataset. Ensemble classifiers such as AdaBoost1 are often used to improve accuracy of weak learners (J. Chandrasekaran.,. 2020). (Refer to appendix section 1 Table 2 for PCA results)

Next, we proceeded to vary the amount of variance captured starting from 0.85 in increments of 0.5 to 0.99 to investigate the trend of variance captured. (Refer to appendix Table 3 full results)

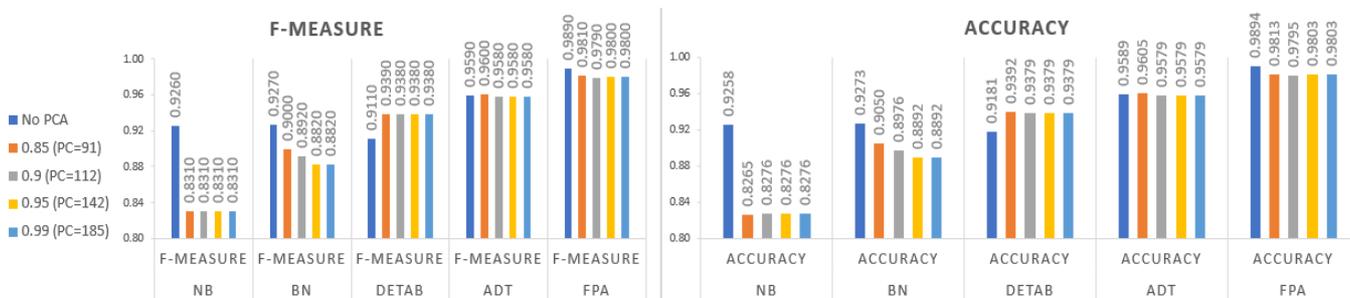

**Figure 6: F-measure and Accuracy graph at different explained variance**

At a variance of 0.85, both NB and BN suffered a drop in accuracy of 0.099 and 0.022. This indicates that some of the important features may have been lost during dimension reduction (Durmuş et al.,. 2021). Varying the amount of variance explained from 0.85 to 0.99 exhibit negligible differences of about 0.001 accuracy for most algorithms. DETAB accuracy and F-measure increased by 0.021 and 0.028 at PCA 0.85. Further increase in PCA from 0.85 to 0.99 did not contribute to any improvement. To conclude, increasing explained variance from 0.85 to 0.99 in general does not bring about significant improvements. NB does not perform well with PCA dataset in all explained variance. (Refer to appendix Section 1 graph 1 for full results)

The Malgenome dataset has class imbalance based on the proportion of benign and malware APKs. The imbalance ratio (IR) of benign and malware is 2.015. Synthetic Minority Oversampling Technique known as SMOTE is often used to oversample the minority class (Durmuş Özkan Şahin et al.,. (2021), Chen, Zhenxiang et al.,. (2017)) to eliminate class imbalance issue.

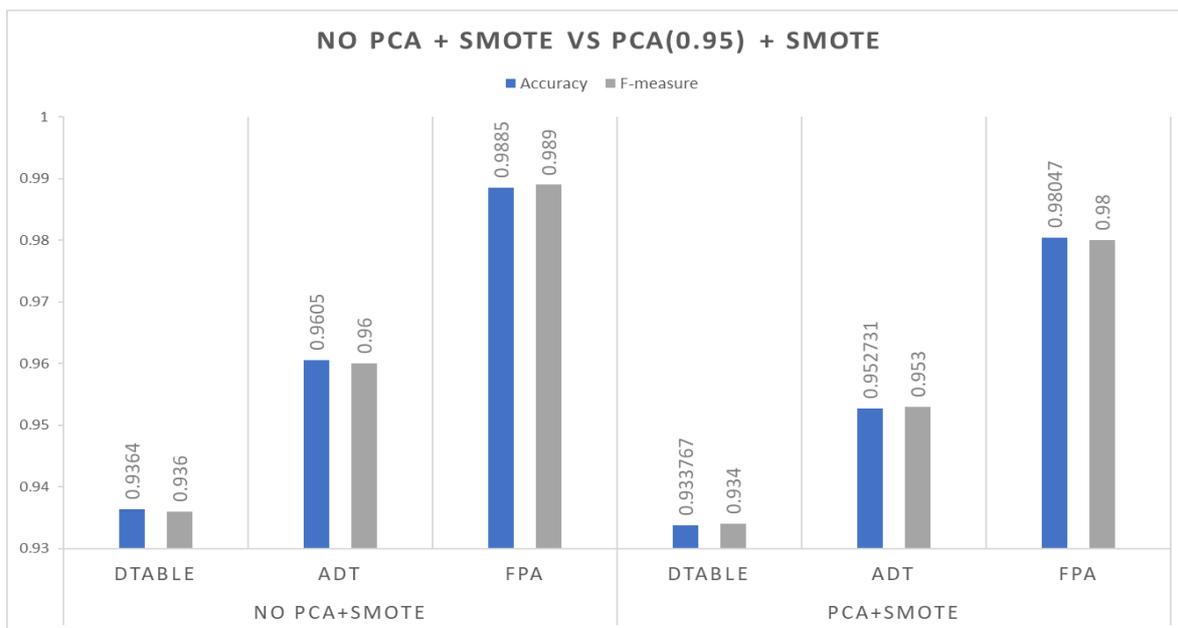

**Figure 7: Accuracy and F-measure results for PCA dataset with and without SMOTE**

Based on the data preprocessing details in the Akintola A.G. experiment, we randomized the dataset prior to the 70:30 split. SMOTE was applied to rebalance the training dataset, which will be used for 10-fold cross validation. Like Akintola A.G. findings, SMOTE improved the overall accuracy, F-measure, and AUC of most algorithms, with FPA also had a 31.25% improvement in FPR. Next, we performed PCA on the Malgenome dataset. We then performed a 70:30 split on the PCA dataset prior to applying SMOTE on training dataset. Again, both NB and BN suffered a drop in accuracy while both DS and CR improved in accuracy (Refer to appendix Section 1 for Table 4 and graph 2 for entire results). These results and behaviors were similar to applying PCA on the original dataset without performing SMOTE. PCA did not further improve the accuracy of SMOTE results.

In the second literature, Abdullah T.A. conducted a study on Android malware detection with six supervised ML classifiers. Two evaluation methods were used in his research, namely holdout validation with 80% training, 20% testing dataset and 10-fold cross validation. 10-fold cross validation was based on the mean score of total folds. Similar to Abdullah T.A., we used Jupyter notebook and Python 3.8. Figure 8 shows the parameters and models used in his literature. We managed to reproduce results close to the literature.

| Algorithm | Evaluation Method | Parameters | Algorithm | Evaluation Method | Parameters |
|---|---|---|---|---|---|
| k-NN | Holdout | n_neighbors=1 | SVM | Holdout | kernel='linear' |
| k-NN | 10-Fold | n_neighbors=1 | SVM | 10-Fold | kernel='linear' |
| DT | Holdout | depth=13 | RF | Holdout | n-estimator=33 |
| DT | 10-Fold | depth=14 | RF | 10-Fold | n-estimator=66 |

Figure 8: Parameters and Models used for HOLDOUT and 10-FOLD

In this section, we compared the performance of classifiers based on Accuracy and F1-score. (Refer to appendix Section 1 Table 4 for test results). For PCA, we set the explained variance to 0.95.

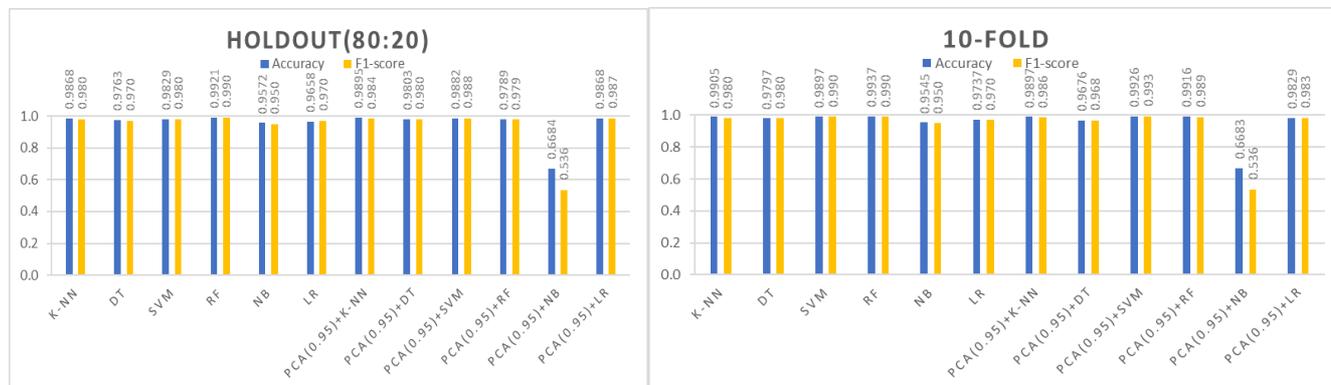

Figure 9: No PCA and PCA Accuracy and F-measure results for HOLDOUT & 10-FOLD

Based on figure 9, the accuracy of k-NN in holdout increased by 0.003 while 10-Fold resulted in a drop by 0.01. The F1-score saw improvement for k-NN by 0.004 while 10-Fold improved by 0.006. Decision Tree had accuracy improvement in Holdout by 0.004 and a decrease in accuracy by 0.012 for 10-Fold. The F-measure also improved for Holdout by 0.04 but decreases by 0.012 in 10-Fold. Both SVM and LR saw an improvement in PCA results for both accuracy and F1-score. For SVM, accuracy increased by 0.005 for Hold out and 0.003 for 10-fold. For F1-score, it increased by 0.008 and 0.003. For LR, accuracy increased by 0.021 for Hold out and 0.009 for 10-fold. For F1-score, it increased by 0.017 and 0.013. In general, PCA successfully transformed and reduced the number of features in the Malgenome dataset from 215 to 142 components at 0.95 explained variance, while giving good results and even about 2% improvements in SVM and LR (bilinear). However, NB algorithms had the worst performance. PCA generated negative correlation values while centering which will result in error in NB using the Multinomial model. We set to MinMaxScalar() from StandScaler() to normalize the input values to 0 and 1. The result drops significantly for NB using the Multinomial model for F1-score by 0.414 and Accuracy by 0.289 for Holdout, 0.286 and 0.414 for 10-Fold. Our AUC result for NB at 0.5 indicates that the NB model is not useful. (Refer to appendix Section 1 Table 6)

Next, we proceeded to vary the explained variance from 0.85 in increments of 0.05 to 0.99 to investigate performance of Classifiers as explained variance increases. Most algorithms performed well based on Accuracy and F1-score despite PCA reducing the number of components. We removed NB from the overall graph as the model exhibits low performance in PCA.

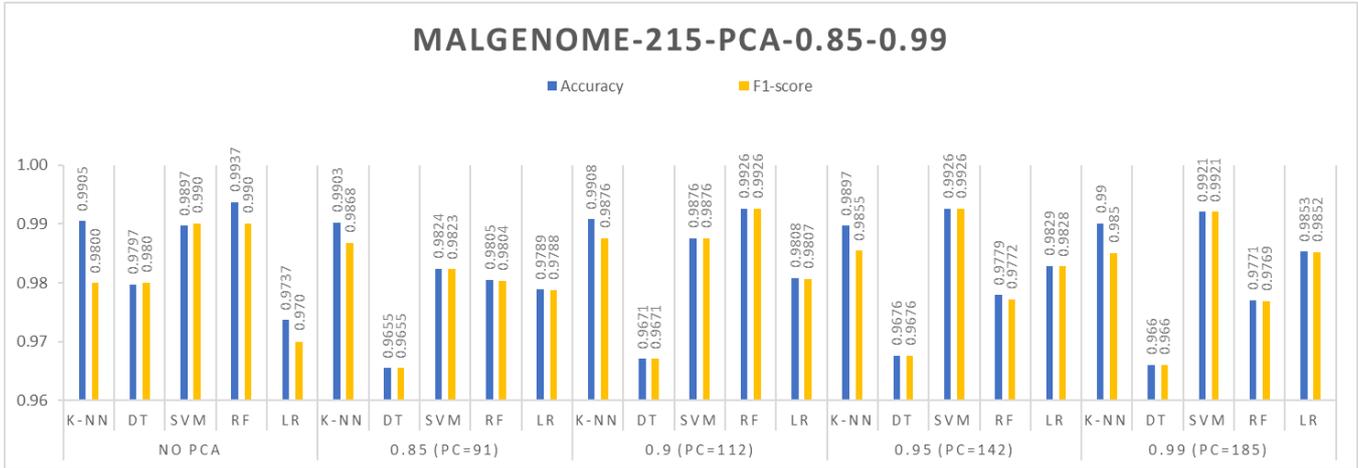

Figure 10: Accuracy and F1-score Results Based on Different PCA explained variance

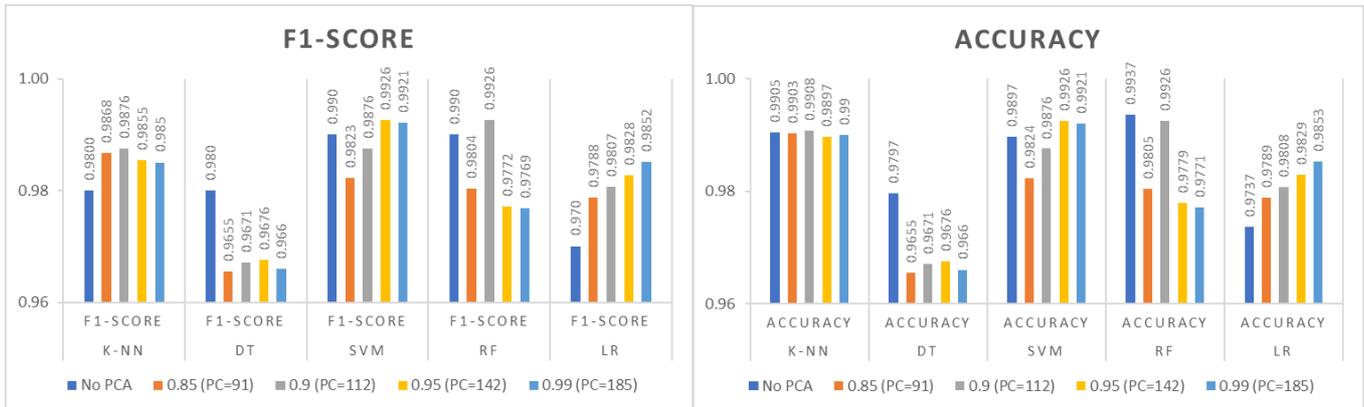

Figure 11: Comparison of Accuracy and F1-score Results Based on Different PCA explained variance

From figure 11, for k-NN, PCA improved the F1-score to 0.9876 with explained variance set to 0.90. Further increases in explained variance saw decrease with the F1-score. DT had the best F1-score at 0.98 without PCA. The best F1-score with PCA for DT was 0.9676 at explained variance of 0.95. For SVM, F1-score at explained variance of 0.85 decreased to 0.9823 and then improved when explained variance was increased. The peak F1-score was 0.9926. For RF, PCA at 0.85 initially decreased the F1-score to 0.9804 and then it peaked at 0.9926 when PCA was 0.90. Further increase of explained variance results in decreased performance. LR (bilinear) shows an increase in F1-score as the explained variance increased. The peak F1-score occurred at explained variance of PCA at 0.99.

Based on the Accuracy graph, the trend of incremental increase of explained variance from 0.85 to 0,95 was identical to F1-score for most classifiers apart from k-NN, where PCA did not have much impact on the accuracy. DT saw a decrease in accuracy of 0.0142 with PCA dataset. Further increase in explained variance to 0.95 only saw an improvement of 0.0021. Due to the imbalance nature of the Malgenome dataset, F1-score, which balances between Precision and Recall value is a more suitable evaluation metric compared to accuracy.

Based on our findings from both research for Malgenome dataset, PCA successfully reduced the number of features from 215 to 98 at explained variance of 0.85, which is a 54.4% reduction in features, while continuing to achieve good results for FPA, ADT, k-NN, SVM, DETAB and LR algorithms. However Naive Baye algorithm is not suitable for PCA dataset as AUC, area under receiver operating characteristic curve (ROC) is 0.5, like tossing a coin. While application of SMOTE on unbalanced dataset did improve the Accuracy of NB (2.68%), BN(1.77%), CR(5.98%), DETAB(1.83%), ADT(0.13%), DS(4.23%) and FPA(-0.09%), application of PCA to Malgenome dataset, follow by SMOTE did not improve the accuracy and f-measure.

## 6.2    Results From Drebin-215

For Drebin dataset, we chose 2 literatures. With Suleiman Y. Yerima 's research, they studied the 5 machine learning models and proposed DriodFushion framework (S. Y. Yerima and S. Sezer, et al.,2019). Their ML results were compared to the results from the DriodFushion Framework. For our study, we focused on applying PCA to the same 5 models and studied the results. To get accurate comparisons, we tried to achieve equivalent results by following the parameters mentioned in the research and using the same machine learning tool which is WEKA. We used 10-fold cross-validation to validate the results.

We set 4 different variance R=0.85, R=0.9, R=0.95 and R=0.99 for PCA to compare the effect of variance on datasets and models. The evaluation metrics of Precision M, Recall M, Precision B, Recall B, Weight F Measures were used for this study.

| Classifier | PrecM | RecM | PrecB | RecB | W-FM |
|---|---|---|---|---|---|
| J48 | 0.972 | 0.964 | 0.979 | 0.984 | 0.9766 |
| REPTree | 0.976 | 0.951 | 0.972 | 0.986 | 0.9730 |
| Random Tree-100 | 0.975 | 0.978 | 0.987 | 0.985 | 0.9824 |
| Random Tree-9 | 0.947 | 0.971 | 0.983 | 0.968 | 0.9672 |
| Voted Perceptron | 0.969 | 0.950 | 0.971 | 0.982 | 0.9701 |

**Table 2: Results from cited literature**

Based on the results, the effects of different PCA R values were insignificant on J48. The difference was less than 1% and ranged from 0.969 to 0.971. On the other hand, increased in PCA values improved results for Voted Perceptron classifier. When PCA's R value was 0.85, weight F measure of Voted Perceptron was 0.964 but when R value was 0.99, the weighted F measure increased to 0.973. Other evaluation metrics also increased slightly when R value was increased.

As for REPTree classifier, the best results were obtained when R value was 0.85. When R value was increased by 0.05, the results dropped slightly but the results improved again when R value was 0.95. However, when R value was increased to 0.99, the results declined and became lower than the results when R value was 0.85. These fluctuations of results showed that over fitting or under fitting of features will not get the optimum results.

As for Random Tree, two different models were used. We had assumed that when author mentioned Random Tree 100 and Random Tree 9, 100 and 9 are the parameters for K value. Random Tree-100 study illustrated that the best PCA R value for this model is 0.95. When R value of 0.90 and 0.99 were applied, the result metrics were almost the same and lowest among all 4 different R value experiments. The second-best values have resulted when R value was 0.85. On second experiment for Random Tree, K value was changed to 9. During this experiment, the results were remarkedly declined when higher R values were applied. R=0.85 gave the best results and R=0.99 gave the lowest results.

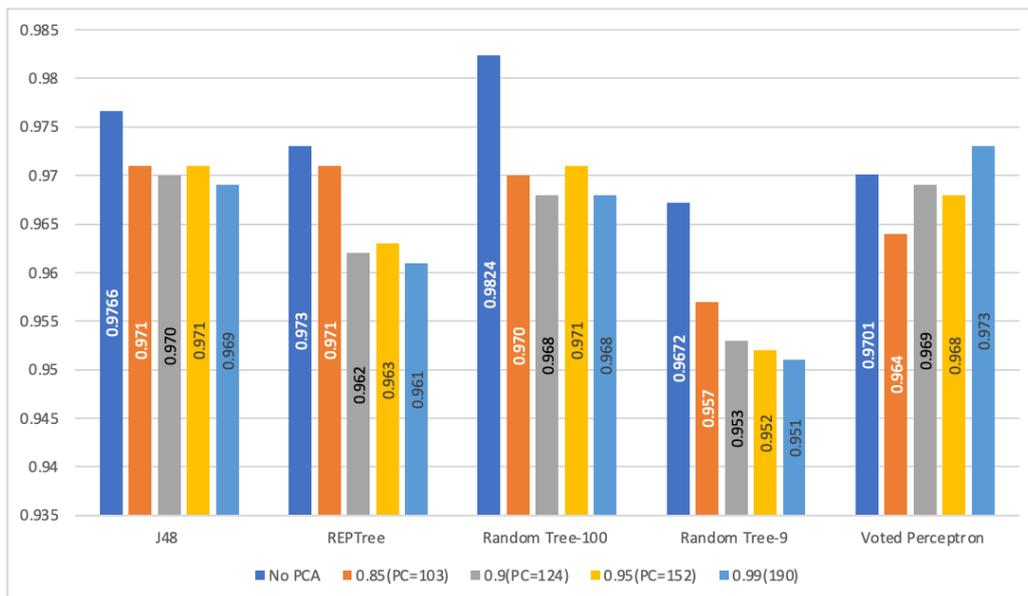

Figure 12: Weighted F-Measure Results

Except for the Voted Perceptron Classification, the rest of the models gave better results without PCA. As for Voted Perceptron, when PCA R value of 0.99 was applied, the Weight FM measure was slightly better than original results. Additionally, all 4 experiments showed that Precision and Recall value of Benign were higher than Precision and Recall value of Malware. The difference was more significant in Random Tree-9 model.

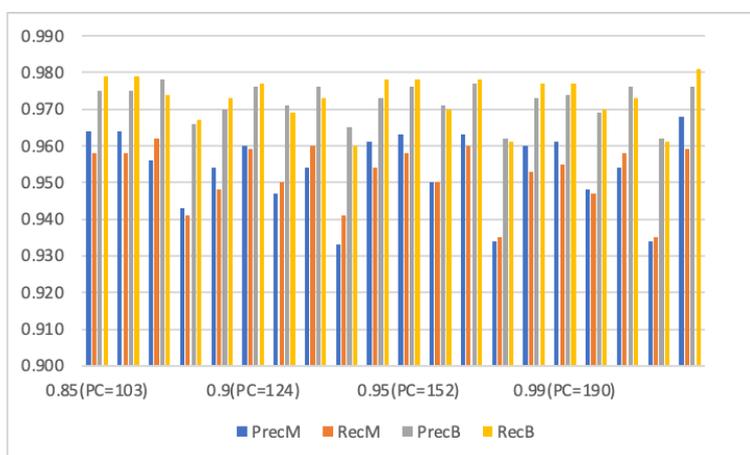

Figure 13: Precision and Recall Results Based on Different PCA R Values

The second dataset we had chosen was "Droid-NNet: Deep Learning Neural Network for Android Malware Detection" published in IEEE International Conference on Big Data (Big Data), 2019. This article purposed the deep neural network called "Droid-NNet" to detect malware. "Driod-NNet" was specifically modelled to detect Android datasets. The literature also studied three traditional classification models such as Decision Tree, Support Vector Machine (SVM) and Logistic Regression. The results were compared to illustrate the robustness and effectiveness of Droid-NNet. (M. Masum and H. Shahriar, et., 2019)

However, for our study, we have chosen results from traditional classification models to apply PCA.

Python 3.8 and scikit-learn library were used to collect the evaluation results. The results were validated by using 10-fold cross-validation and using a standard scalar to scale the dataset. We also reused the same parameters used in original paper. 'rbf' for SVM Kernel vale, 'Gini' for Decision Tree criterion, and L2 for the Logistic Regression penalty. The evaluation metrics utilized are True Positive Rate(TPR), False Positive Rate and F-beta score .

| Classifier | TPR (True Positive Rate) | FPR (False Positive Rate) | F beta Score |
| --- | --- | --- | --- |
| Decision Tree | 0.973810 | 0.019305 | 0.978411 |
| SVM | 0.961111 | 0.008280 | 0.981564 |
| Logistic Regression | 0.976190 | 0.004850 | 0.988858 |

**Table 3: Results from cited literature**

The Drebin dataset has an unbalanced ratio of data and accuracy of the results may not give accurate representation of model's performance. Thus, F-beta score was used to determine the performance of the models. (M. Masum and H. Shahriar, et.,2019). To give more weight to recall, beta value 10 was used. We applied 4 values for R: 0.85, 0.90,0.95,0.99. However, none of the R values gave better results compared to original F-beta Score.

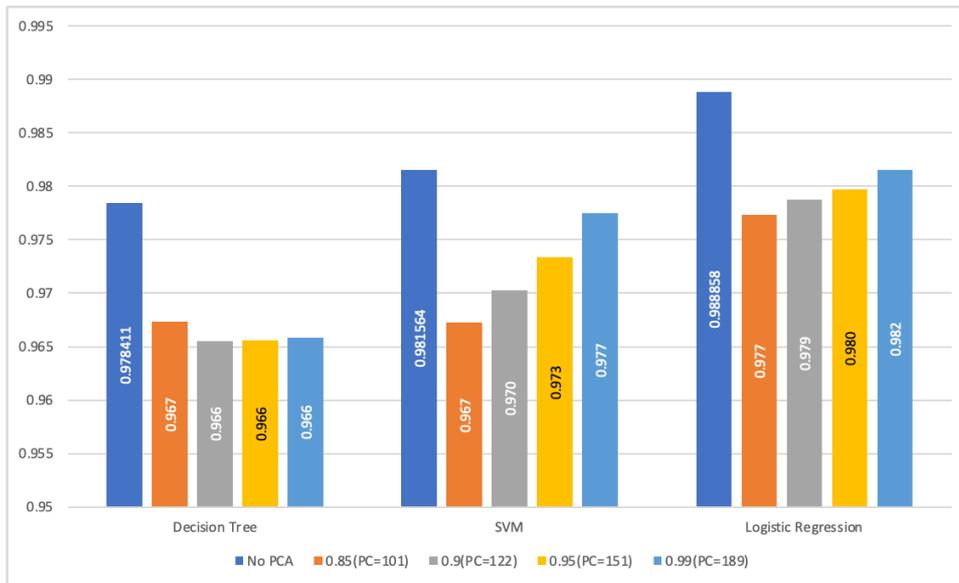

**Figure 14: F-beta Score Results Based on Different PCA R Values**

For Decision Tree, the results declined after PCA was applied. When PCA 0.85 was applied, the F-beta score value dropped to 0.967 from 0.978. R value of 0.90,0.95 and 0.99 gave the same results at 0.966. SVM also resulted with poorer results compared to the original results. At R=0.85, F-beta score was decreased by 1%. However, the value improved gradually when we applied higher R values. When R=0.99 was applied, the difference was only 0.004.

Similarly, Logistic Regression gave lower results in 0.85 but the results were slightly improved when R=0.90 was applied. These results showed us that applying PCA does not necessarily improve F-beta Score. As for the other evaluation metric, please refer to the appendix.

## 6.3 Results from Cic-invesandmal2019

For the CIC-InvesAndMal2019 dataset, we have selected 2 reference papers for our research. First Paper, (Sangal, Aviral., 2020) applied the Principal Component Analysis (PCA) which is a feature reduction technique for malware detection. For the data processing phase, the researchers ensured that there is no missing value. Principal Component Analysis was applied after data processing phase and total of 100 attributes was selected. Cross-validation 10-fold was being applied for the classification. However, there was no mention on the variance applied for PCA. For the classifier, Naive Bayes (NB), Support Vector Machine (SVM), Random Forest, Decision Tree, and Nearest Neighbours (K1) were used in the paper. The researcher performed the experiment by WEKA, which we similarly applied and performed in our research. The accuracy of the related paper is as shown below in Table 4.

| Classifier PCA | Accuracy Result | Precision | Recall | F-Measure |
|---|---|---|---|---|
| Naive Bayes (NB) | 88.23% | 0.877 | 0.882 | 0.877 |
| SVM | 91.26% | 0.912 | 0.913 | 0.908 |
| Random Forest | 96.05% | 0.96 | 0.961 | 0.969 |
| Decision Tree (J48) | 92.90% | 0.929 | 0.929 | 0.931 |
| Nearest Neighbours (ibk) K 1 | 93.88% | 0.939 | 0.939 | 0.925 |

Table 4: Results from sited Literature above, Aviral Sangal [5]

Second Paper, (Viraj Kudtarkar., 2020) had only 384 samples of botnet applications and 1105 samples of clean non-malicious applications. The researcher extracted the data using APK tool after the decompression of the APK files and extraction of the required features from the source code. This extraction included essential information such as intents and user permissions. A total of 18 features were selected in data selection and data-pre-processing phase. The researcher then divided the data into 70:30 portions for training and testing respectively. For the classification algorithms, five classifiers were used for training which are Naive Bayes (NB), Support Vector Machine (SVM), Random Forest, Decision Tree, and Logistic Regression as below in Table 5. The result from the cited paper is as shown below without PCA.

| Classifier | Accuracy Result | Precision | Recall | F-Measure |
|---|---|---|---|---|
| Naive Bayes | 94.80% | 0.841 | 0.958 | 0.891 |
| SVM | 83.10% | 0.841 | 0.958 | 0.891 |
| Random Forest | 87.60% | 0.879 | 0.97 | 0.917 |
| Decision Tree | 94.30% | 0.945 | 0.982 | 0.959 |
| Logistic Regression | 95.40% | 0.949 | 0.994 | 0.964 |

Table 5: Results from sited Literature above, (Viraj Kudtarkar.,. 2020)

Both papers had 4 common algorithms with one additional that differed from the other. For our experiment, we included both additional algorithms namely Nearest Neighbours and Logistic Regression to align measurements. We defined 4 different PCA variance R=0.85, R=0.9, R=0.95 and R=0.99 to compare which has the most effective and reliable model. Accuracy, Precision, Recall and F Measures are the evaluation metric for the algorithm. Random Forest had the highest accuracy after the application of PCA with 96.05%. Naïve Bayes had the lowest accuracy rate of 88.23%. Meanwhile, the second paper had the highest accuracy result of 95.40% for Logistic Regression classifier. On the other hand, SVM had low accuracy compared to others with 83.10%. For Decision Tree, Navie Bayes, and Logistic Regression, after applying PCA with different variance saw their accuracies drop significantly.

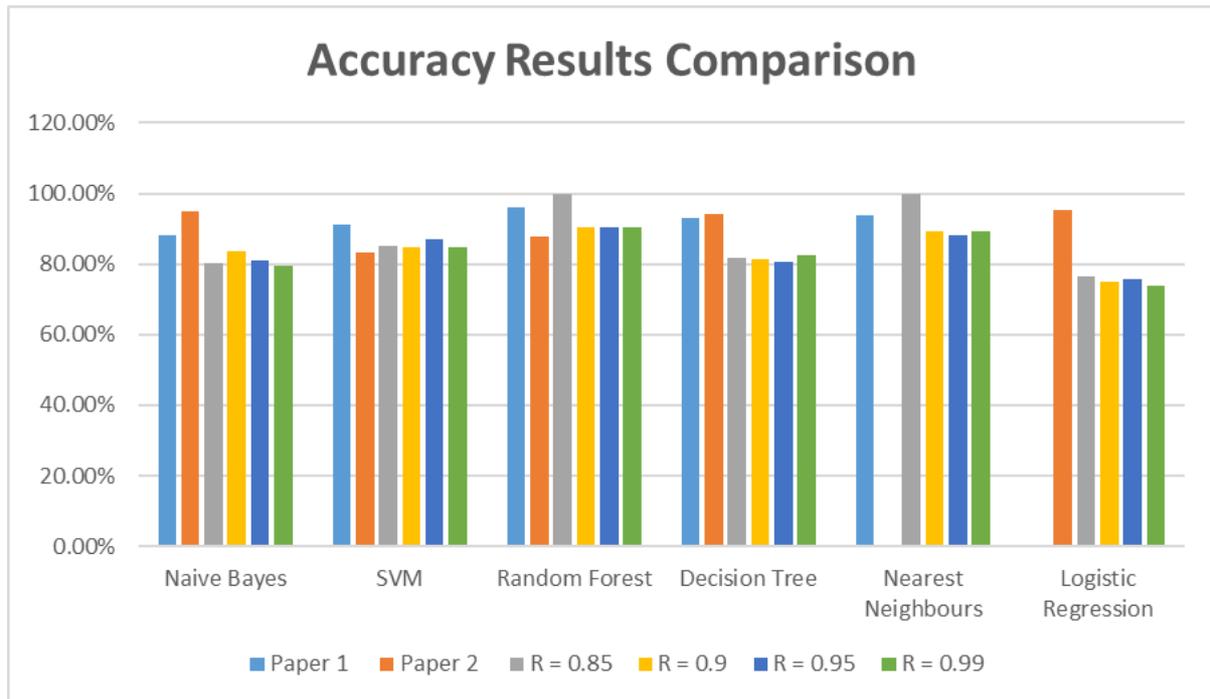

**Figure 15: Accuracy Results Based on Different PCA R Values**

Based on our results, the variance change R did not significantly change for the accuracy except for the Random Forest while using R=0.85 which resulted in the best accuracy of 99.89%. Other results are 90.40% and 90.30% for Random Forest while using different variances, the rest of the classifiers had significant changes in accuracy depending on the variance applied. From the result we observed that variance R=0.85 is the best for Random Forest and K Nearest Neighbor classifiers which had the best accuracy. When R value was increased slightly by 0.05 and set to 0.9, the results dropped significantly on Random Forest and K nearest Neighbors. There was no significant difference between 0.95 and 0.99. The recall result gradually decreased when the R value is increased. The recall result as of 0.85 has "0.871" and it's gradually dropping as "0.741,0.742,0.736". This proved that overfitting or underfitting of features will not achieve optimum results.

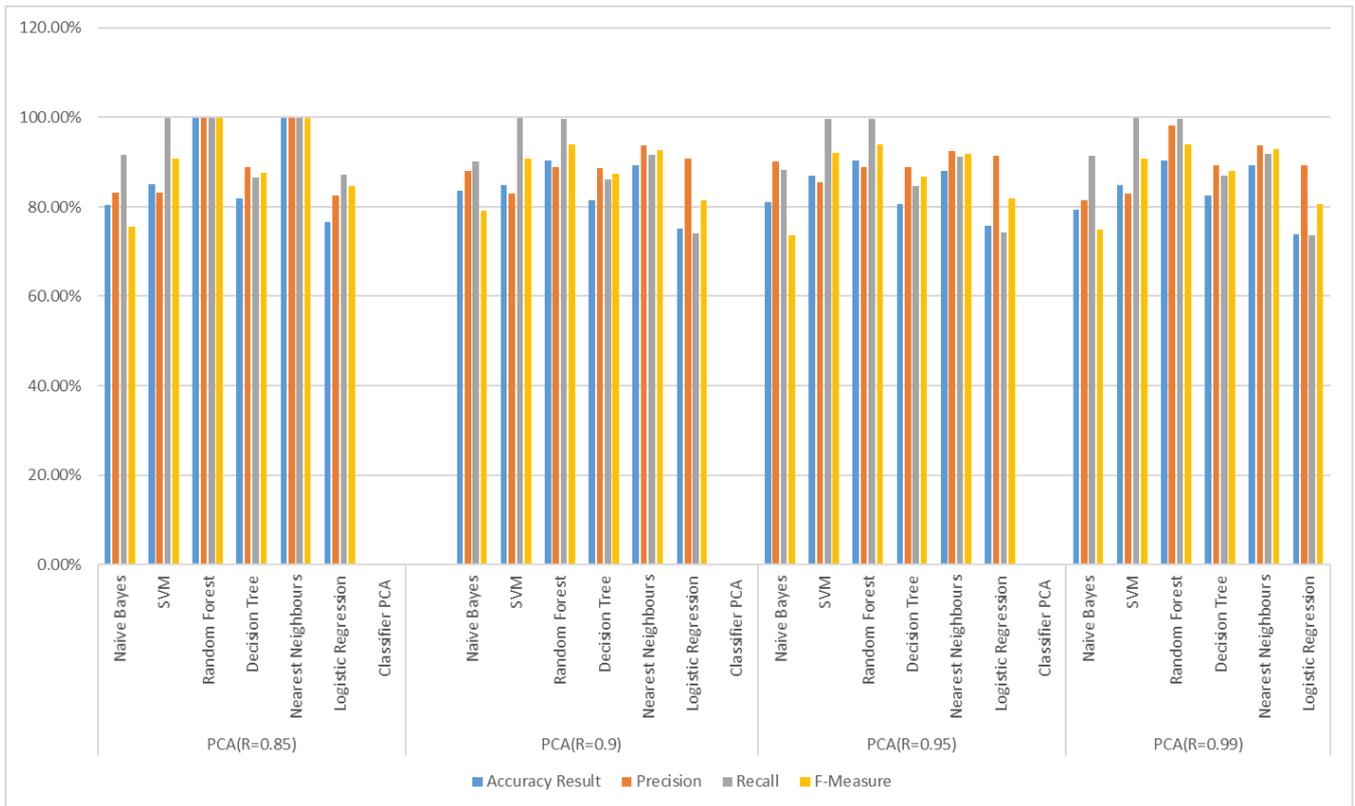

**Figure 16: Accuracy, Precision, Recall & F-Measure Results Based on Different PCA R Values**

## 8. CONCLUSION & DISCUSSION

Principle Component Analysis (PCA) is a dimension reduction technique widely used in various scientific industries that handles large high dimensional dataset, where the number of features is greater than the amount of data. For example, they are commonly used in computer vision, image recognition and to produced 2-dimensional data for visualization. However, there has been a lack of study in the effects of dimensional reduction technique to Computational Forensics and specifically with malware detection and analysis. This study focuses on the application of PCA in the analysis of 3 different malware analysis related datasets containing Android malware. We also evaluate the impact of imbalance classes within the dataset with dimensional reduction.

Based on our experiments where we conducted parallel experiments done by other researchers on the same datasets, we observed that the performance measurements of machine learning models generally comparable with marginal degradation with dimensional reduced datasets after the application of PCA. We did observe notable degradation with the reduction of variance range of 0.85 to 0.99 when PCA is applied to the datasets. There were a few improvements observed with no noted pattern or specific algorithms. With imbalance classes within the dataset, we observed that the application of Synthetic minority oversampling technique (SMOTE) then with applied dimensionality reduction had performance degrade with the reduction of variance.

For future works, we hope to extent our work to different malware datasets which includes dynamic and memory forensic dataset for Android malware. Other Dimension reduction and feature selection techniques can be included in our future studies to improve processing of high dimensional data generated by intrusion detection systems.